\documentclass[12pt]{article}

\usepackage{layout}
\usepackage[textwidth=460pt, textheight=630pt,voffset=15pt]{geometry}
\usepackage{fancyhdr}
\usepackage{comment}
\usepackage{changepage}
\usepackage{tocloft}
\usepackage[para]{footmisc}
\usepackage{soul}

\fancypagestyle{noheader}{
  \fancyhf{}
}

\usepackage[T1]{fontenc}
\usepackage[title]{appendix}%
\usepackage{textcomp}%
\usepackage[dvipsnames]{xcolor}%
\usepackage{multicol}
\usepackage{enumitem}
\usepackage{etoolbox}

\usepackage{multirow}%
\usepackage{amsmath,amssymb,amsfonts}%
\usepackage{amsthm}%
\usepackage{mathrsfs}%
\usepackage{mathtools}
\usepackage{stmaryrd} 
\usepackage{accents} 
\usepackage{tensor}

\usepackage{graphicx, tikz}%
\usepackage{subcaption}
\usetikzlibrary{cd,graphs,decorations.pathmorphing,decorations.markings}
\usepackage{tabularx}
\usepackage{wrapfig}

\usepackage[framemethod=TikZ]{mdframed} 
\mdfsetup{%
   middlelinecolor=BurntOrange, 
   middlelinewidth=.8pt,
   backgroundcolor=white, 
   roundcorner=10pt}

\usepackage{url}
\usepackage[pdftex,colorlinks=true, pdfstartview=FitV, linkcolor= Mlink, citecolor=Mlink, urlcolor= Mlink, hyperindex=true, hyperfigures=true, breaklinks=true]{hyperref}

\usepackage{authblk}
\usepackage[numbers,sort&compress]{natbib}

\newlength{\bibitemsep}
\newlength{\bibparskip}
\let\oldthebibliography\thebibliography
\renewcommand\thebibliography[1]{%
  \oldthebibliography{#1}%
  \setlength{\parskip}{\bibitemsep}%
  \setlength{\itemsep}{\bibparskip}%
}
 \usepackage{etoolbox}
 \newcommand\addtag[1][]{%
   \refstepcounter{equation}\hfill(\theequation)%
   \notblank{#1}{\label{#1}}{}}

\input{aastexv6.sty}
\input{Short_cut_3.0.sty}

\title{\vspace{-2.2cm}\Large\bf A natural model for curved inflation}

\begin{document}

\author[1,2]{{Quentin} {Vigneron}\footnote{\href{mailto:quentin.vigneron@umk.pl}{quentin.vigneron@umk.pl}}}
\author[3]{{Julien} {Larena}\footnote{\href{mailto:julien.larena@umontpellier.fr}{julien.larena@umontpellier.fr}}}
\affil[1]{\small\it{Institute of Astronomy, Faculty of Physics, Astronomy and Informatics}, {Nicolaus Copernicus University}, {{Grudzi{\k{a}}dzka 5}, {Toru\'n}, {87-100} {Poland}}}
\affil[2]{\small\it{ENS de Lyon, CRAL UMR5574, Universit\'e Claude Bernard Lyon 1, CNRS, Lyon, F-69007, France}}
\affil[3]{\small\it{Laboratoire Univers \& Particules de Montpellier, Universit\'e de Montpellier, CNRS}, {Montpellier}, {France}}


\maketitle


\vspace{-.7cm}
\begin{abstract}


{Inflationary models with a non-zero background curvature {require additional hypothesis or parameters compared to flat inflation and the procedure to construct them cannot be as simple as in the flat case.} For this reason, there is no consensus on the primordial power spectrum that should be considered at large scales in a curved Universe. 
In this letter, we propose a model of curved inflation {in which the usual canonical quantization and Bunch--Davies vacuum choice of the flat case can be considered.} The framework is a recently proposed modification of general relativity in which a non-dynamical topological term is added to the Einstein equation. 
{The model is universal as it is the same for any background curvature, and no additional parameters or hypothesis on the initial spatial curvature are introduced.} This gives a natural and {simple} solution to the problem of {constructing} curved inflation, and at the same time provides an additional argument for this topological modification of general relativity.}

\end{abstract}

\vspace{.6cm}

Cosmological inflation successfully explains the Cosmic Microwave Background (CMB) anisotropies, and provides at the same time a solution to the flatness and horizon problems. The success of the model lies in its simplicity: single field slow-roll inflation with a Bunch-Davies vacuum condition allowing for a satisfactory fit of the CMB data \citep{2020_Planck_X}. However, while inflation in general relativity (GR) predicts a negligible late-time spatial curvature, it is {a priori} not necessarily negligible at the start of inflation, and can even be required to be non-zero by the topology of the Universe. Therefore, for the consistency of the theory, one should be able to easily construct an inflationary scenario regardless of the initial curvature. There is however no consensus on the construction of such a scenario in GR, and in particular on the shape of the power spectrum that should result from it. {All scenarios in the literature require additional parameters, fine-tuning, or different mechanisms than the usual Bunch--Davies vacuum choice, adding considerable complexity compared to the flat case} \citep[e.g.][]{1995_Lyth_et_al, 1998_Garriga_et_al, 2002_Gratton_et_al, 2005_del_Campo_et_al, 2005_Lasenby_et_al, 2016_Bonga_et_al, 2017_Ratra, 2019_Akama_et_al, 2019_Handley, 2021_Thavanesan_et_al, 2021_Renevey_et_al, 2022_Ratra, 2022_Hergt_et_al, 2022_Renevey_et_al, 2023_Letey_et_al, 2023_Dagostino_et_al}. 
The difficulty is mainly due to the fact that, for non-zero background curvature $K$, the evolution equation for the curvature perturbation $\CR$ is no longer the solution of a wave equation, and features a non-local operator $\CZ$ \citep{2019_Handley}:

\begin{align}\label{eq_MS_K}
	&v_K'' - \left(\frac{\CZ''}{\CZ}  + 2K + \frac{2K\CZ'}{\CH\CZ} + \Delta\right)v_K = 0, \\
    &{\rm with} \ \ v_K \coloneqq \CZ\CR \quad ; \quad \CZ \coloneqq z\sqrt{\frac{\CD^2}{\CD^2-K\CE}}, \nonumber 
\end{align}
where $\CH$ is the conformal Hubble rate, $\CD^2 \coloneqq \Delta + 3K$ and $\CE\coloneqq \frac{\varphi'^2}{2\CH^2}$ with $\varphi$ the background inflaton. This makes it impossible to use standard quantization procedures to define initial conditions and predict the primordial power spectrum, leading to the use of non-standard initial conditions (e.g.~kinetic dominance \cite{2014_Handley_et_al, 2022_Hergt_et_al, 2023_Letey_et_al}), {or different mechanisms (e.g. the one-bubble inflation for negatively curved background \cite{1993_Sasaki_et_al, 1995_Yamamoto_et_al, 1996_Yamamoto_et_al, 1997_Sasaki_et_al, 1998_Garriga_et_al})}. Additionally, $\CR$ can only be conserved on super-Hubble scales during inflation if the curvature scale is also super-Hubble, as shown by
\begin{align}
	\frac{1}{\CH}\CR' &= \frac{1}{\CE} \left\{\frac{-k^2+3K}{\CH^2}\Psi + \frac{K}{\CH^2}\left[\frac{\Psi'}{\CH} + \Psi\right]\right\}, \label{eq_R_prime_K_RG}
\end{align}
where $\Psi$ is the Bardeen potential and $k$ the wave number. Without this hypothesis, numerical resolution is necessary to track the evolution of the power spectrum \citep{2019_Handley, 2022_Renevey_et_al}. 
Clearly, it is quite unsavory to require a priori that spatial curvature is super-Hubble simply to ensure conservation of $\CR$, since this is supposed to be one of the main achievements of inflation. 
It is also possible to work with a variable solution of a wave equation without non-local operator, as shown by~\cite{2023_Letey_et_al}. However, there is still not a natural initial condition that can be considered, the super-Hubble conservation is not fulfilled, and a numerical resolution is necessary.

Additionally, at the level of the background, changes in the evolution of the scale factor induced by the presence of curvature (e.g. the bounce present in the spherical case, or the upper bound on the Hubble radius in the hyperbolic case) further complicate the case for curved inflation in GR.
In particular, additional fine-tuning on the number of e-folds is required \citep{2003_Uzan_et_al, 2021_Renevey_et_al, 2022_Hergt_et_al}, and, for $K>0$, different power spectra are obtained if the initial conditions are taken before or after the bounce \citep{2022_Renevey_et_al}.

Overall, the procedure to obtain a power spectrum resulting from an inflationary phase in a curved universe in GR is considerably more difficult than that of the flat case. In this sense, with a background curvature, we lose the simplicity and naturalness of the inflation hypothesis that made its success.

The goal of this letter is to propose a curved inflationary scenario bypassing the problems previously mentioned, as simple as flat inflation, and without introducing any additional parameters. The philosophy of our approach lies in a slight modification of the Einstein equation that was proposed by \citep{2024_Vigneron}, rather than within GR. This slight, albeit fundamental, deviation is amply compensated by the resulting simplicity of the model and the fact that the initial motivation for the modification is not linked at all to inflation.

We first proceed with the presentation of this modification of GR, which we name topo-GR, along with its perturbation equations that were derived in \cite{2023_Vigneron_et_al_b}. All the related equations from this theory are presented and compared with the ones from GR in Table~\ref{tab_Summary}. We then show how curved inflation is constructed in this theory, in particular through the use of a modified curvature perturbation variable, and give the resulting power spectra. Details of the derivations can be found in the Appendices. Throughout this letter, we use Greek letters for spacetime indices, and Roman letters for spatial indices.\\

\textbf{\large Topo-GR}\\
\begin{figure}
    \centering
    \includegraphics[width=.70\columnwidth]{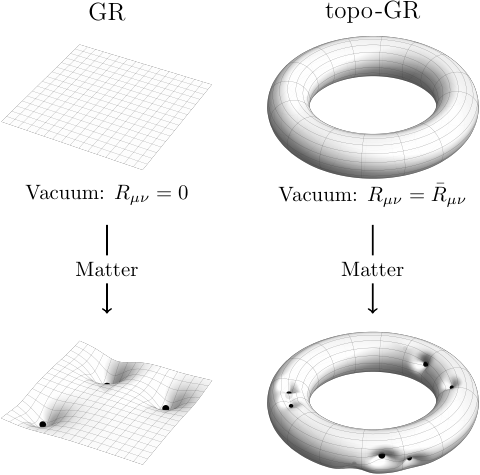}
    \caption{\small Qualitative difference on the role of matter between GR (left) and topo-GR (right). In GR, spacetime is Ricci flat in vacuum, i.e. $R_{\mu\nu} = 0$, and matter induces a deviation from flatness. In topo-GR, in vacuum, spacetime is curved by topology, i.e. $R_{\mu\nu} = \bar R_{\mu\nu}$, and matter induces a deviation from that `topological curvature'. \label{fig_diff}}
 \end{figure}
The original motivation for the modification of GR proposed in \cite{2024_Vigneron} is not related to inflation, but is to obtain a relativistic field equation which admits a non-relativistic limit in any topology, something not possible in GR (see Sec. 4 in \citep{2024_Vigneron}). This modification does not add any degrees of freedom compared to GR, and the equivalence principle is preserved. Physical differences only appear at scales comparable to the topology scale. In this theory, the spacetime Ricci curvature $R_{\mu\nu}$ in the Einstein equation is replaced by the difference between that tensor and a reference Ricci curvature $\bar R_{\mu\nu}$ that is fixed and depends on the spatial covering space of the spatial topology, i.e. $R_{\mu\nu} \rightarrow R_{\mu\nu} - \bar R_{\mu\nu}$, leading to equation~\eqref{eq_topo_GR_eq_1}, in Table~\ref{tab_Summary}. {We direct the reader to \cite{1995_La_Lu, 2024_Vigneron} for the precise definitions of the topological notions involved in the definition of the topo-GR theory, involving in particular the Thurston classification of closed 3-manifolds}. The additional term appearing in the Einstein equation is conserved by the second equation~\eqref{eq_topo_GR_eq_2}.


The precise way $\bar R_{\mu\nu}$ is chosen as function of the topology is described in Sec. 5.2.2 of \cite{2024_Vigneron}. In the present paper, where we consider perturbations around a homogeneous and isotropic background, $\bar R_{\mu\nu}$ takes the form~\citep{2023_Vigneron_et_al_b}
\begin{align} \label{eq_gauge_Rbar}
    \bar R_{\mu\nu} = 2K\left(\gamma_{\mu\nu} + \CL_{X}\gamma_{\mu\nu}\right),
\end{align}
where $X^\mu$ is a first order vector field, and $\gamma_{\mu\nu}$ is the background comoving spatial metric. For $K$ = 0, we have $\bar R_{\mu\nu} = 0$, and topo-GR is equivalent to GR.

Qualitatively, within topo-GR, matter does not directly curve spacetime, as in GR, but only induces deviations (not necessarily small) from the curvature ``imposed'' by topology, i.e. deviations from~$\bar R_{\mu\nu}$. This is seen through the modified Einstein equation~\eqref{eq_topo_GR_eq_1} from which, in the absence of matter, $R_{\mu\nu}$ reduces to the reference (topological) curvature $\bar R_{\mu\nu}$ (see also Fig.~\ref{fig_diff}).\\

\textbf{\large Perturbation equations}\\
Table~\ref{tab_Summary} summarises the perturbation equations around the homogeneous and isotropic solution of topo-GR that were derived in \citep{2023_Vigneron_et_al_b} (right column), and compare them with GR (left column), using the longitudinal gauge, i.e.
\begin{align}
    g_{\mu\nu} \dd x^\mu \dd x^\nu &= a^2\left\{-(1+2\Phi)\dd\tau^2 - Q_i\dd x^i \dd \tau + \left[(1-2\Psi)\gamma_{ij} + h_{ij}\right]\dd x^i \dd x^j\right\}.\label{eq_gauge}
\end{align}
The two systems of equations are equivalent for $K=0$, and differ only for $K\not=0$. Consequently, flat inflation is equivalent in GR and topo-GR. For $K\not=0$, three main differences of topo-GR compared to GR are particularly relevant to the design of  inflationary scenarios.

First, the background equations~\eqref{eq_Blind_Back} are the flat Friedmann equations for any curvature, i.e. the scale factor evolution is independent of the value of $K$. {This implies that the dynamics of the scale factor, and also of the background inflaton, are equivalent to the dynamics of these variables in GR for flat background.} Especially, during a de Sitter phase, we have
\begin{align}
	a(t) \propto -\frac{1}{\tau} \quad\quad {\rm for} \quad\quad \tau<0, \quad \quad \forall \, K. \label{eq_back_blind}
\end{align}
This is a radical difference with respect to GR since there is no more bounce or upper-bound for the Hubble radius when $K\not=0$. Consequently, we will be able to use the limit $\tau\rightarrow -\infty$ to obtain an exact Bunch-Davies vacuum for any curvature.

\begin{table*}[t!]
\centering
\footnotesize
\caption{\small Comparison between GR and topo-GR for the homogeneous and isotropic solution and its first order perturbations (longitudinal gauge) \cite{2023_Vigneron_et_al_b}. Terms in \colblue{\bf{bold blue}} are present in GR but missing (or different) in topo-GR. Terms in \colred{\bf{bold red}} are present in topo-GR but missing (or different) in GR. For $K=0$, these two systems are equivalent. We defined $\kappa \coloneqq 8\pi G$.}
\label{tab_Summary}
\centering
\renewcommand{\arraystretch}{1.25}
\newcommand{\vspacetable}[1]{\multicolumn{2}{c}{} \vspace{#1}\\}
\begin{tabular}{p{\columnwidth/2 - 25pt}|p{\columnwidth/2}}
	\hline\hline 
	\centering \bf General relativity & \centering \bf topo-GR \arraybackslash\\
     \hline 
    \multicolumn{2}{c}{\sc Field equations} \vspace{.1cm}\\
    $R_{\mu\nu} = \kappa \left(T_{\mu\nu} - \frac{T_{\alpha\beta}g^{\alpha\beta}}{2} \, g_{\mu\nu}\right)$
    &
    $R_{\mu\nu} - \colred{\boldsymbol{\bar R_{\mu\nu}}} = \kappa \left(T_{\mu\nu} - \frac{T_{\alpha\beta}g^{\alpha\beta}}{2} \, g_{\mu\nu}\right)$\addtag[eq_topo_GR_eq_1]\\
    
    \centering$\emptyset$
    &
    $\colred{\boldsymbol{\nabla^\nu\left(\bar{R}_{\mu\nu} - \frac{\bar{R}_{\alpha\beta}g^{\alpha\beta}}{2} g_{\mu\nu}\right) = 0}}$ \addtag[eq_topo_GR_eq_2]\\
    \vspacetable{-10pt} 
    \hline 
    \multicolumn{2}{c}{\sc Homogeneous and isotropic solution} \\
	$\begin{aligned}
		&\CH^2 = a^2\frac{\kappa}{3}\rho -\colblue{\boldsymbol{K}} \\
		&\CH' = -a^2\frac{\kappa}{6}\left(\rho + 3p\right)
	\end{aligned}$&
	$\begin{aligned}
		&\CH^2 = a^2\frac{\kappa}{3}\rho \\
		&\CH' = -a^2\frac{\kappa}{6}\left(\rho + 3p\right)
	\end{aligned}$ $\quad\quad\colred{\boldsymbol{\forall K}}$ \addtag[eq_Blind_Back]\\
    \vspacetable{-10pt}
	\hline
	\multicolumn{2}{c}{\sc First order perturbations} \\
	\multicolumn{2}{c}{Scalar modes} \\
 	$(\Delta + \colblue{\boldsymbol{3K}})\Psi = a^2\frac{\kappa}{2}\delta\rho_l + 3\CH\left(\Psi' + \CH\Phi\right)$
    &
	$\Delta\Psi = a^2\frac{\kappa}{2} \delta\rho_l  + 3\CH\left(\Psi' + \CH\Phi\right) +\colred{\boldsymbol{K\Delta\CC}}$ \addtag[eq_blind_pert_S_1] \\

 	$\Psi' + \CH \Phi = -a^2\frac{\kappa}{2}(\rho+p)\CV$
    &
	$\Psi' + \CH \Phi = -a^2\frac{\kappa}{2}(\rho+p)\CV + \colred{\boldsymbol{K\CC'}}$ \addtag[eq_blind_pert_S_2]\\

  	$\Psi - \Phi = a^2\kappa\Pi$
    &
	$\Psi - \Phi = a^2\kappa\Pi + \colred{\boldsymbol{4K\CC}}$ \addtag[eq_blind_pert_S_3] \\

    $\begin{aligned}
		&\Psi'' + 2\CH\Psi' + \CH\Phi' + (2\CH' + \CH^2)\Phi -\colblue{\boldsymbol{K\Psi}} = \\
			&\qquad a^2\frac{\kappa}{2}\left(c_{\rm s}^2\delta\rho_l + \delta p_{\not= \rm ad} + \frac{2}{3}\Delta\Pi\right) \\
	\end{aligned}$
    &
	$\begin{aligned}
	    &\Psi'' + 2\CH\Psi' + \CH\Phi' + (2\CH' + \CH^2)\Phi = &\\
			&\qquad a^2\frac{\kappa}{2}\left(c_{\rm s}^2\delta\rho_l + \delta p_{\not= \rm ad} + \frac{2}{3}\Delta\Pi\right) + \colred{\boldsymbol{K\Delta\CC}} 
   \end{aligned}$ \addtag[eq_blind_pert_S_4]\\
   
	\multicolumn{2}{c}{Vector modes} \\
    $\left(\Delta + \colblue{\boldsymbol{2K}}\right)Q_i = -a^2 2\kappa (\rho + p)(\CV_i - Q_i)$
    &
	$\left(\Delta - \colred{\boldsymbol{2K}}\right)Q_i = -a^2 2\kappa (\rho+p)\left(\CV_i - Q_i\right) 
                + \colred{\boldsymbol{4K\CC_i'}}$ \addtag[eq_blind_pert_V_1]\\

	$Q_i' + 2\CH Q_i = a^2\kappa\Pi_i$
    &
    $Q_i' + 2\CH Q_i
            = a^2\kappa\Pi_i + \colred{\boldsymbol{4K\CC_i}}$ \addtag[eq_blind_pert_V_2]\\
	\multicolumn{2}{c}{Tensor mode}\\
	$h_{ij}'' + 2\CH h_{ij}' + \left({\colblue{\boldsymbol{2K}} - \Delta}\right)h_{ij} = 2a^2\kappa \Pi_{ij}$ &
	$h_{ij}'' + 2\CH h_{ij}' + \left({\colred{\boldsymbol{6K}} - \Delta}\right)h_{ij} = 2a^2\kappa \Pi_{ij}$\addtag[eq_blind_pert_T]\\
	\multicolumn{2}{c}{Conservation of $\bar{R}_{\alpha\beta} - \frac{\bar{R}_{\mu\nu}g^{\mu\nu}}{2} g_{\alpha\beta}$}\\
	\centering$\emptyset$
	&
    $\colred{\boldsymbol{\CC'' + 2\CH\CC' - \Delta\CC }}\colred{\boldsymbol{= -a^2\kappa \Pi}}$\addtag[eq_blind_pert_C_1] \\ 
 	\centering$\emptyset$
    &
	$\colred{\boldsymbol{\CC_i'' + 2\CH\CC_i' + \left(2K -\Delta\right) \CC_i}} \colred{\boldsymbol{= -a^2\kappa \Pi_i}}$\addtag[eq_blind_pert_C_2]\\ 
    \vspacetable{-10pt}
	\hline\hline
\end{tabular}
\end{table*}

Second, the scalar mode equations~\eqref{eq_blind_pert_S_1}--\eqref{eq_blind_pert_S_4} no longer feature $K\Psi$. That term is one of the reasons a non-local operator appears in GR, because it requires inverting the Laplacian in the Poisson equation to eliminate $\Psi$ when deriving the equation for $\CR$. {Its} absence in the topo-GR equations will allow us to obtain a wave equation without non-local operator.\\
\indent Third, additional gauge invariant variables are present: a scalar mode $\CC$ and a vector mode $\CC_i$, both being solutions of wave equations sourced by the matter anisotropic stress [equations~\eqref{eq_blind_pert_C_1} and~\eqref{eq_blind_pert_C_2}]. These variables come from the first order contribution to $\bar R_{\mu\nu}$ in the gauge~\eqref{eq_gauge}, and can be related to the tilt of the reference observer induced by $\bar R_{\mu\nu}$ in that gauge \cite{2023_Vigneron_et_al_b}. For a vanishing stress, equation
~\eqref{eq_blind_pert_C_1} implies that $\CC$ decays during radiation and matter dominated areas on sub-Hubble scales, but is conserved on super-Hubble scales. Hence if generated during inflation, $\CC$ could have an impact on the power spectrum at large scales. For this reason, we will consider $\CC\not=0$. The same can be said for $\CC_i$ with equation~\eqref{eq_blind_pert_C_2}. However, while this variable can be non-zero, even if it sources the metric vector mode $Q_i$ through equations~\eqref{eq_blind_pert_V_1} and~\eqref{eq_blind_pert_V_2}, these imply that $Q_i$ always decays on all scales for vanishing stress, and therefore can be neglected. Since observational imprints of vector modes come from $Q_i$, we can also neglect $\CC_i$. This shows that vector modes are not important in topo-GR, as in GR. Consequently, from now on we will not consider these modes anymore.\\

\textbf{\large Curvature deviation perturbation}\\
We define the  {\it{curvature deviation perturbation}} as 
\begin{align}
    \tilde\CR \coloneqq \CR - K\CC.
\end{align}
Physically, $\tilde\CR$ is related to the difference between the first order spatial scalar curvature $\delta\left({^{(3)}R}\right)$ and the first order spatial reference curvature $\delta\left(\gamma^{ij} \bar R_{ij}\right)$ in the comoving gauge (with subscript $_c$):
\begin{align}
	 \left[\delta\left({\CR^{(3)}}\right) - \delta\left(g^{ij} \bar R_{ij}\right)\right]_{\rm c}
		&= \frac{4}{a^2}\Delta \tilde\CR. \label{eq_diff_R3_Rbar}
\end{align}
For a flat background, $\tilde\CR$ reduces to $\CR$. But while in GR, $\CR$ is the natural variable to use to study inflation for $K=0$, in topo-GR this becomes $\tilde\CR$ for any~$K$. 
There are three main reasons for this:
\begin{enumerate}[leftmargin=*]
    \item $\tilde\CR$ is conserved on super-Hubble scales for shear-free adiabatic perturbations or single field inflation, without assuming additionally the curvature scale to be super-Hubble. This results from
    \begin{align}
	   &\CH\tilde\CR' =  \frac{a^2\kappa}{3(1+w)}\left[\cs^2\delta\rho_{\rm c} + \delta p_{\not= ad} + \frac{2}{3}\left(\Delta + 3K\right)\Pi \right]. \label{eq_tilde_R_prime_main}
    \end{align}
    For adiabatic perturbations, that formula also shows that the density perturbation after inflation is directly generated by $\tilde\CR$ for any $K$, as with $\CR$ in GR for $K=0$.
    \item With a single scalar field as a source, $\tilde\CR$ is solution of the Mukhanov-Sasaki equation for any $K$:
    \begin{align} \label{eq_wave_tilde_R}
	   &\tilde\CR'' + \frac{2z'}{z}\tilde\CR' - \Delta \tilde\CR = 0 \qquad \forall K,
    \end{align}
    where $z \coloneqq \frac{a\varphi'}{\CH}$. This equation for $\tilde \CR$ is formally equivalent to that of $\CR$ in GR for a flat background, but with the Laplacian depending on spatial curvature.
    \item $\tilde\CR$ is one of the two free scalar fields to appear in the action at second order (see next section).
\end{enumerate}
The fact that $\tilde\CR$ is the relevant variable to consider in topo-GR rather than $\CR$ is consistent with the main idea of the theory: what matters for the dynamics is not the (spatial/spacetime) curvature, but the difference between that curvature and the reference curvature.\\

\textbf{\large Second order action}\\
To summarize, single field inflation in topo-GR features three independent variables: two scalar modes $\tilde\CR$ and $\CC$, and a tensor mode $h_{ij}$, solutions of the wave equations~\eqref{eq_wave_tilde_R}, \eqref{eq_blind_pert_C_1} and \eqref{eq_blind_pert_T}, respectively. Additionally, the scale factor evolves as $a \propto -1/\tau$ for any curvature during an exact de Sitter phase. Therefore, apart from the additional field $\CC$, the situation is formally equivalent to that of GR with a flat background. Thus, a canonical quantization with a Bunch-Davies vacuum choice can be implemented. For this, the second order action is necessary, allowing us to define the canonical variables.

{The action for topo-GR is \citep{2024_Vigneron}\footnote{{Note that a first order action involving the distorsion tensor, i.e. the difference between the Levi-Civita connection of the metric and the reference connection, can also be written for topo-GR \citep{2024_Vigneron}.}}:}
\begin{align}\label{eq_Action_topo_GR}
    S = \int \dd x^4 \sqrt{|g|}\left[ \frac{m_{\rm p}^2}{2}\left(R_{\mu\nu} - \bar R_{\mu\nu}\right) g^{\mu\nu} + \CL_{\rm \varphi}\right],
\end{align}
where $m_{\rm p}$ is the Planck mass and $\mathcal{L}_\varphi$ the inflaton Lagrangian. The second order expansion of this action can be expressed as
\begin{align}\label{eq_Action_Second_topo}
S = \frac{1}{2}\int \dd\tau \dd^3x \sqrt{\gamma} &\left\{\left[\left(\tilde v'\right)^2 + \frac{z''}{z} \tilde v ^2- \left(D_i \tilde v \right)^2 \right] \right.  \nonumber \\
    &\quad- \left[C' \frac{\CD^2}{K} C' + \frac{a''}{a} C \frac{\CD^2}{K} C - D_i C D^i \frac{\CD^2}{K} C\right]  \\
	&\quad \left. + \left[\left(\mathfrak h'_{ij}\right) ^2 + \frac{a''}{a} \left(\mathfrak h_{ij}\right) ^2 - \left(D_k \mathfrak h_{ij}\right)^2 - 6K \left(\mathfrak h_{ij}\right)^2\right] \right\}, \nonumber 
\end{align}
where we introduced $\tilde v \coloneqq a m_{\rm p} \sqrt{2\epsilon} \, \tilde\CR$, $C \coloneqq a m_{\rm p} K \sqrt{2} \, \CC$ and $\mathfrak{h}_{ij} \coloneqq \frac{a m_{\rm p}}{2} h_{ij}$, and with $\epsilon \coloneqq 1-\CH'/\CH$. This action can be compared to the much more complicated second order action in GR which, for $K\not=0$, features a non-local operator, as derived in \citep{2002_Gratton_et_al, 2019_Handley}. 
This further emphasizes the simplicity of the equations obtained when topo-GR is considered over GR once $K\not=0$. 

The Lagrangian for $C$ is fourth order in the spatial derivatives due to the presence of the operator ${\CD^2 \coloneqq \Delta + 3K}$. However, the resulting field equation is still the wave equation~\eqref{eq_blind_pert_C_1}, but with that operator applied on it. Additionally, the presence of $\CD^2$ implies that oscillating modes for $C$ such that ${\CD^2 = -k^2 + 3K >0}$ have a negative kinetic term and $C$ is a ghost field for those modes. However, due to the discretization of $k$ for $K>0$, only $k = 0$ is permitted among those modes, which is pure gauge and not physical. The second discrete mode $k^2 = 3K$ is also not physical as it disappears from the action. This can also be seen from the perturbation equations which can be rewritten to depend only on $\CD^2 \CC$ (Sec.~4.4 in \cite{2023_Vigneron_et_al_b}). Hence, the physical modes of $C$ always have $-k^2 + 3K <0$ and~$C$ is a canonical field. Consequently, the presence of $\CD^2$ in the action will only affect the amplitude of the primordial power spectrum of $K\CC$.\\

\textbf{\large Choice of initial conditions}\\
For a de Sitter phase, the background evolution~\eqref{eq_back_blind} implies that, for any $K$, we have $a''/a \rightarrow 0$ and $z''/z \rightarrow 0$ in the limit $\tau\rightarrow-\infty$, and the variables $\tilde v$, $C$ and $\mathfrak{h}_{ij}$ become harmonic oscillators in a static background (almost for $C$). That background can be described as a ``curved Minkowski spacetime'', i.e. homogeneous and isotropic without expansion, but spatially curved. Consequently, in the limit $\tau \rightarrow -\infty$, a canonical quantization can be applied on these three variables along with a Bunch-Davies choice, leading to
\begin{align} \label{eq_IC_vtilde}
    \tilde v(\tau, k)           &\overset{\tau\rightarrow - \infty}{\sim} \frac{e^{-ik\tau}}{\sqrt{2 k}}, \\
    C(\tau, k)                  &\overset{\tau\rightarrow - \infty}{\sim} {\frac{K}{k^2 - 3K}} \frac{e^{-ik\tau}}{\sqrt{2 k}} && \ \  k^2 >3 K, \label{eq_IC_C} \\
    \mathfrak{h}_{ij}(\tau, k)  &\overset{\tau\rightarrow - \infty}{\sim} \frac{e^{-i\sqrt{k^2+6K}\tau}}{\sqrt{2 \sqrt{k^2+6K}}} &&\ \ k^2 + 6K > 0. \label{eq_IC_h}
\end{align}
The frequency $\sqrt{k^2+6K}$, instead of $k$, appearing for the tensor mode results from its effective mass being~$k^2 +6K$. Additionally, the extra factor of {$K/(k^2 - 3K)$} appearing in the normalisation of $C$ comes from its canonical momentum being {$\pi_C = -\frac{\CD^2}{K} C'$}, i.e. different from the usual formula $\pi_C = C'$. This normalisation is well-defined since only modes with $k^2 > 3K$ exist for $C$, as explained before.\\

\textbf{\large Power spectra from slow-roll inflation}\\
Using the wave equations~\eqref{eq_blind_pert_T}, \eqref{eq_blind_pert_C_1} and \eqref{eq_wave_tilde_R}, the scale factor evolution~\eqref{eq_back_blind} and the initial conditions~\eqref{eq_IC_vtilde}--\eqref{eq_IC_h}, we can derive the power spectra of $\tilde\CR$, $\CC$ and $h_{ij}$ from slow-roll inflation in the same way it is done in GR with a flat background. The only difference is that the wave number $k$ refers to harmonic functions adapted to the given spatial curvature.

Defining the power spectrum as $\langle \Psi_{\bf k}(t)\Psi_{\bf k'}(t)\rangle = (2\pi)^3\delta(\bf k+ k') \CP_\Psi$, and  $\CP_\Psi \coloneqq |\Psi_k(\tau)|^2$, the primordial power spectra resulting from a slow-roll inflationary phase are 
\begin{align}
	\CP_{\tilde \CR}	&= \frac{1}{\epsilon} \frac{H^2}{2 m_{\rm p}^2} 2^{2\nu-4} \left(\frac{\Gamma(\nu)}{\Gamma(\frac{3}{2})}\right)^2 \, \frac{k^{-6\epsilon + 2\eta}}{k^3}, \label{eq_PtildeR}\\
	\CP_{K\CC}			&= \frac{H^2}{2 m_{\rm p}^2} 2^{2\mu-4}  \left(\frac{\Gamma(\mu)}{\Gamma(\frac{3}{2})}\right)^2\, \left(\frac{K}{k^2 - 3K}\right)^2\,  \frac{k^{-2\epsilon}}{k^3}, \\
	\CP_h			  &= 16 \frac{H^2}{2m_{\rm p}^2} 2^{2\mu-4} \left(\frac{\Gamma(\mu)}{\Gamma(\frac{3}{2})}\right)^2 \, \frac{\left(\sqrt{k^2+6K}\right)^{-2\epsilon}}{\left(\sqrt{k^2+6K}\right)^{3}},
\end{align}
where $H = \CH/a$ is the Hubble rate during inflation, and we have defined the usual slow-roll parameters $\epsilon \coloneqq - \dot H/H^2$ and $\eta \coloneqq \epsilon - \ddot\varphi/(H\dot\varphi)$; moreover, we set $\mu\coloneqq 3/2+\epsilon$ and $\nu \coloneqq 3/2 + 3\epsilon - \eta$.

This is is an unambiguous way, free from new parameters, of defining the scaling and amplitude of a spectrum resulting from a single field slow-roll inflation in any curved space, something out of reach in GR. The power spectrum of the additional scalar mode $\CC$ is fully fixed by the slow-roll parameters and the curvature, and no additional parameters are needed. Initial conditions on the density perturbation are obtained from $\CP_{\tilde\CR}$ using relation~\eqref{eq_tilde_R_prime_main}.

At first order in the slow-roll parameters, we have the following ratios \begin{align}
     \CP_{K\CC} / \CP_{\tilde\CR}   &= \epsilon {\left(\frac{K}{k^2 - 3K}\right)^2}, \\
     \CP_{h} / \CP_{\tilde\CR}     &= 16\epsilon \left(\frac{k}{\sqrt{k^2 + 6K}}\right)^3.
\end{align}
The tensor-to-scalar ratio $r$ features a scale dependence for scales approaching the curvature scale, with an enhancement for $K<0$, similarly to the GR case \cite{2023_Dagostino_et_al}. While there is a divergence for $k^2 = -6K$, it might be avoided for specific closed hyperbolic topologies for which the minimal wave number of tensor modes would be $k_{\rm min}^2 > -6K$ (e.g. \cite{1999_Cornish_et_al}). Additionally, the consistency relation between $r$ and the scaling index $n_h = -2\epsilon$ of tensor modes features a dependence on the wave number for $K\not=0$. This prediction would be observable only for a current curvature scale comparable to the CMB scale.

But most importantly, while topo-GR features one additional gauge invariant variable~$\CC$
, the amplitude of its power spectrum is negligible compared to that of the curvature deviation perturbation. Additionally, $\CC$ decays during radiation and matter dominations on sub-Hubble scales. Therefore, its effects on scalar mode fluctuations can be neglected from the start of inflation to the late time Universe, and considering $\CC = 0$ is a good approximation. In this case $\tilde\CR$ exactly corresponds to the usual curvature perturbation $\CR$, and the situation is even closer to that of GR for $K=0$. Especially, the scale dependence and the amplitude of $\CR$ is equivalent for any $K$. {This is the main difference between our model and other curved inflationary scenarios in which $\CP_{\CR}$ depends on $K$ (see e.g. \citep{2019_Handley})

For this reason, observations will only be able to distinguish between the inflationary model presented in this work (and therefore the topo-GR theory) and current curved inflationary models in GR if $\Omega_K$ is large enough. Currently, $\Omega_K$ is constrained to be of the order of $10^{-3}$ or less within GR \citep{2020_Planck_X} or topo-GR \citep{2023_Vigneron_et_al_b}, and future surveys such as the Euclid space telescope might allow us to gain one order of magnitude on this constraint \citep{2022_EUCLID_prep_XV}. Since this leads to a very large curvature scale, this might already be too small due to the cosmic variance present at large scales. If so, this model of inflation would still have a theoretical interest, using Occam's Razor argument, as no additional parameters or hypothesis on the initial curvature are needed compared to curved inflationary models in GR to incorporate curvature into inflation. Post-inflationary constraints would then have to be used to probe topo-GR. With this regard, studying the dynamics of reheating might be an interesting perspective. The main difference we expect with reheating scenarios in GR would come from the presence of the second reference connection $\bar\nabla$ in topo-GR. This offers the possibility of reheating scenarios based on a coupling between the inflaton and that connection, as has been studied within the broader context of Metric Affine Gravity theories \citep[e.g.][]{2019_Shimada_et_al}. We let this study for a future work.}\\

\textbf{\large Conclusion}\\
Curved inflationary scenarios are essential to the coherence of the inflation hypothesis, but cannot be easily constructed within general relativity. 
In this letter, {we have proposed a model of curved inflation which keeps the simplicity of flat inflation while allowing for canonical quantization with any background spatial curvature. For this, a non-dynamical topological term is added to the Einstein equation.}
{The main features of the model are the following}: (i) no bounce and no upper bound on the Hubble radius exists during inflation and a Bunch-Davis vacuum condition is well-defined for any $K$; (ii) there exists a variable $\tilde \CR$, reducing to the curvature perturbation for $K=0$, that is conserved on super-Hubble scales and is solution of the Mukhanov-Sasaki equation for any $K$; (iii) a unique well-defined primordial power spectrum is obtained for any $K$, with the same scale dependence for scalar modes in each case. 

This model of curved inflation has the same number of degrees of freedom as in the
Standard Model, 
{and is universal in the sense it is the same regardless of the background spatial curvature, and this with the same procedure than in the flat GR case. Since one of the successes of (flat) inflation is its simplicity, the fact that our model manages to keep this simplicity for any curvature is a strength}, which gives a natural and simple answer to how curved inflation should be built.

We advocate that this is also a strong argument for considering a non-dynamical topological term in the Einstein equation in the form proposed by the topo-GR theory, when non-asymptotically flat spacetimes and non-Euclidean topologies are considered.\\

\noindent\textbf{Acknowledgments:} QV is supported by the Centre of Excellence in Astrophysics and Astrochemistry of Nicolaus Copernicus University in Toru\'n, and by the Polish National Science Centre under grant No. SONATINA 2022/44/C/ST9/00078. 
We thank Hamed Barzegar, Elliot Lynch and Sébastien Renaux-Petel for valuable discussions, and Guillaume Laibe for his comments on the manuscript.

\vspace{-.2cm}

\footnotesize
\bibliographystyle{apsrev4-1}
\bibliography{bib_General}
\normalsize

\appendix


\section{Vector modes during inflation in topo-GR}
\label{app_Vector_modes}

The equations constraining the vector modes $Q_i$ and $\CC_i$ during single field inflation are
\begin{align}
	\left(\Delta - 2K\right)Q_i &= 4K\CC_i' \, , \label{eq_Qi_inf_1}\\
	Q_i' + 2\CH Q_i &=  4K\CC_i. \label{eq_Qi_inf_2}
\end{align}
The wave equation~\eqref{eq_blind_pert_C_2} for $\CC_i$ follows from these two equations. In topo-GR, $Q_i$ is sourced by $\CC_i$, which seems to indicate that vector modes might be non-negligible, for $K\not=0$. 
However, during a de Sitter phase, characterised by $a(\tau) \propto 1/\tau$ for any $K$, the general solution of this system of equations is
\begin{align}
    \CC_i   &= \CC_{i, 1} \left[e^{-i \tau \sqrt{k^2+2K}}\left(1 + i\tau\sqrt{k^2+2K}\right) + i e^{i \tau \sqrt{k^2+2K}}\left(1 - i\tau\sqrt{k^2+2K}\right) \right], \\
    Q_i     &= -4 K \CC_{i, 1} \, \tau \left[e^{-i\tau \sqrt{k^2+2K}} + i e^{i \tau \sqrt{k^2+2K}}\right], \label{eq_sol_Qi}
\end{align}
where $\CC_{i, 1}$ is a constant. Therefore, the metric vector mode $Q_i$ decays on all scales and can be neglected.

\section{Conservation of \texorpdfstring{$\tilde\CR$}{}}
\label{app_Conservation}

The combination
\begin{align*}
    \left\{\CH\eqref{eq_blind_pert_S_2} - \frac{2\left[\eqref{eq_blind_pert_S_2}' - \eqref{eq_blind_pert_S_4} + 2\CH\eqref{eq_blind_pert_S_2} + \eqref{eq_blind_pert_C_1}\right]}{3(1+w)}\right\}
\end{align*}
of the first order equations of topo-GR leads to
\begin{align}
	&\CH\tilde\CR' =  \frac{a^2\kappa}{3(1+w)}\left[\cs^2\delta\rho_{\rm c} + \delta p_{\not= ad} + \frac{2}{3}\left(\Delta + 3K\right)\Pi \right], \label{eq_tilde_R_prime}
\end{align}
where $\delta\rho_{\rm C} \coloneqq \delta\rho_\sigma - 3\CH(1+w)\CV$ is the density perturbation in the comoving gauge. Apart for the term $3K\Pi$, that equation is equivalent to the one obtained in GR for the curvature perturbation $\CR$ for $K=0$. Then, using the Poisson equation~\eqref{eq_blind_pert_S_1}, equation~\eqref{eq_tilde_R_prime} leads to 
\begin{align}
	\frac{1}{\CH}\tilde\CR' &\underset{\delta p_{\not=\rm ad}=0}{\overset{\Pi=0}{=}} \quad  \frac{2\cs^2}{3(1+w)}\frac{a^2\kappa}{2\CH^2}\delta\rho_{\rm c} \nonumber \\
		&\quad =\frac{2\cs^2}{3(1+w)}\left[\frac{-k^2}{\CH^2}\left(\Psi - K\CC\right) - 3K\frac{\CC'}{\CH}\right], \label{eq_tilde_R_prime_ad}
\end{align}
for shear-free adiabatic perturbations, and to
\begin{align}
	\frac{1}{\CH}\tilde\CR' &\overset{\varphi}{=} \frac{2}{3(1+w)}\frac{a^2\kappa}{2\CH^2}\delta\rho_{\rm c} \nonumber\\
		&= \frac{1}{\epsilon}\left[\frac{-k^2}{\CH^2}\left(\Psi - K\CC\right) - 3K\frac{\CC'}{\CH}\right], \label{eq_tilde_R_prime_inflation}
\end{align}
for single field inflation.

The wave equation~\eqref{eq_blind_pert_C_1} implies that, for shear-free matter, $\CC$ is conserved on super-Hubble scales during a de Sitter phase, radiation or matter dominated eras. Therefore, the right-hand-sides of equations~\eqref{eq_tilde_R_prime_ad} and \eqref{eq_tilde_R_prime_inflation} vanish on super-Hubble scales, and $\tilde\CR$ is conserved on these scales.

The uniform energy density curvature perturbation $\zeta \coloneqq -\CR - \frac{\CH}{\rho'}\delta\rho_{\rm c}$ is often used in GR to study inflation, since it is also constant on super-Hubble scales for adiabatic matters, or single field inflation, in the flat case. Similarly to~$\tilde\CR$, we can define a uniform energy density curvature deviation perturbation $\tilde \zeta$ as
\begin{align}
	\tilde\zeta	&\coloneqq \zeta + K\CC,
\end{align}
which is also conserved on super-Hubble scales when $\tilde\CR$ is, for any $K$.

\section{Wave equation for \texorpdfstring{$\tilde\CR$}{}}
\label{app_Wave}

Consider a scalar field $\phi$ with energy-momentum tensor
\begin{align}
	T^\mu{}_\nu = \nabla^\mu\phi \nabla_\nu\phi - \frac{1}{2}\nabla^\alpha\phi \nabla_\alpha\phi \, \delta^\mu_\nu - V \, \delta^\mu_\nu.
\end{align}
We define the perturbation of the scalar field in the longitudinal gauge as $\phi = \varphi + \delta\varphi$. The first order matter variables from the scalar field are
\begin{align}
	&\delta\rho_l	= \frac{1}{a^2}\left(\varphi'\delta\varphi' - \varphi'^2\Phi\right) + \partial_\varphi V \delta\varphi \quad ; \quad \CV = -\frac{\delta\varphi}{\varphi'} \quad ;  \quad \Pi = \Pi^i = \Pi^{ij} = 0 \ ; \nonumber\\ 
	&\CV_i = Q_i \quad ; \quad \delta p_{\not=\rm ad} =\left(1-\cs^2\right)\delta\rho_{\rm C}.  
\end{align}
The first order field equations~\eqref{eq_blind_pert_S_1}--\eqref{eq_blind_pert_S_4} of topo-GR become in this case
\begin{align}
    &\Delta\Psi - 3\CH\Psi' - a^2 V\Psi - \frac{1}{2}\left(\varphi'\delta\varphi' + a^2\partial_\varphi V\delta\varphi\right) = K\Delta\CC - 4KV\CC, \label{eq_1} \\
	&\Psi' + \CH \Psi - \frac{1}{2}\varphi'\delta\varphi = 4K\CH\CC + K\CC', \label{eq_2} \\
	&\Phi-\Psi	= - 4K\CC. \label{eq_5}
\end{align}
The derivative of equation~\eqref{eq_2} is
\begin{align}
	&\Psi'' + \CH \Psi' + \CH'\Psi - \frac{1}{2}\left[\varphi'\delta\varphi' - \left(2\CH\varphi' + a^2\partial_\varphi V\right)\delta\varphi\right] = K\left(\CC'' + 4\CH\CC' + 4\CH'\CC\right). \label{eq_3}
\end{align}
Finally, the time component of the conservation of the energy-momentum tensor is, using equation~\eqref{eq_5},
\begin{align}
	&\delta\varphi'' + 2\CH\delta\varphi' + a^2\partial_\varphi^2V\delta\varphi - \Delta\varphi - 4\varphi'\Psi' + 2a^2 \partial_\varphi V \Psi = 4K\left(-\varphi'\CC' + 2a^2\partial_\varphi V \CC\right). \label{eq_4}
\end{align}
The left-hand-sides of equations~\eqref{eq_1}--\eqref{eq_4} are equivalent to the first order equations obtained with the Einstein equation in the flat case, but here for any curvature. On the RHS, additional terms are present that depend on $\CC$. The combination of equations
\begin{align}
    \left\{\eqref{eq_3} + \frac{\CH}{\varphi'}\eqref{eq_4} + \frac{2z'}{z}\eqref{eq_2} - \eqref{eq_1}\right\}, \nonumber 
\end{align}with the use of ${\frac{2z'}{z} = -2\CH - \frac{2\CH'}{\CH} - 2a^2\frac{\partial_\varphi V}{\varphi'}}$ and $z \coloneqq \frac{a\varphi'}{\CH}$, leads to
\begin{align}
	\CR'' + \frac{2z'}{z}\CR' - \Delta \CR = K\CC'' + K\frac{2z'}{z}\CC' - K\Delta\CC.
\end{align}
This is a wave equation for the curvature perturbation, with a source term depending on $\CC$. This equation can be rewritten as,
\begin{align}
	\tilde\CR'' + \frac{2z'}{z}\tilde\CR' - \Delta \tilde\CR = 0 \qquad\qquad \forall\,K,
\end{align}
with $\tilde\CR = \CR - K\CC$.

\section{Second order action}
\label{app_Action}

The action at second order of topo-GR can be calculated following the same approach as in GR, within the ADM formalism \cite{2003_Maldacena}. From the action~\eqref{eq_Action_topo_GR} of topo-GR, we obtain
\begin{align}
	S = \frac{1}{2} \int \dd\tau \dd x^3 \sqrt{s} \Big[ &N  \left(\CR^{(3)}_{ij} - \bar R_{ij}\right) s^{ij} + \frac{1}{N} \left(E^{ij} E_{ij} - E^2\right) \nonumber \\
	&+ \frac{1}{N} \left(\phi' + N^i \partial_i \phi \right)^2 + N D_i \phi D^i \phi - 2N V \nonumber \\
		&\ + \frac{1}{N} N^i N^j \bar R_{ij} - \frac{2}{N} N^i\bar R_{i 0} + \frac{1}{N} \bar R_{00} \Big], \label{eq_Action_ADM}
\end{align}
where $s_{ij}$ is the spatial metric with $\CR^{(3)}_{ij}$ its Ricci curvature and $\sqrt{s} \coloneqq \sqrt{{\rm det}\, s_{ij}}$, $N$ is the lapse, $N^i$ is the shift, and
\begin{align}
	E_{ij} \coloneqq \frac{1}{2} \gamma'_{ij} - D_{(i}N_{j)}, \quad E \coloneqq E^i{}_i.
\end{align}

We first consider only scalar modes. The goal is to develop the action~\eqref{eq_Action_ADM} up to second order in the first order variables $\CR$ and $\CC$. For this, the first and second orders of the reference curvature as functions of these variables are needed. Because the reference curvature is a fixed tensor, its perturbation orders are pure gauge. This means that in a general gauge, up to second order, $\bar R_{\mu\nu}$ is of the form \cite{2023_Vigneron_et_al_b}
\begin{align} \label{eq_gauge_Rbar_big}
    \bar R_{\mu\nu} = \overset{(0)}{R}_{\mu\nu} + \left[\CL_{X}\overset{(0)}{R}_{\mu\nu}\right] + \left[\frac{1}{2}\CL_{X}^2\overset{(0)}{R}_{\mu\nu} + \CL_Y \overset{(0)}{R}_{\mu\nu}\right],
\end{align}
where $X^\mu$ and $Y^\mu$ are respectively first and second order vectors, and $\overset{(0)}{R}_{\mu\nu} = 2K\gamma_{ij}$. Since all second order variables will vanish in the second order action by using the background equations, we are only interested in determining $X^\mu$ in the gauge we are choosing. The time component $X^0$ is not present in the first order of $\bar R_{\mu\nu}$, and in the second order is only present in a spatial boundary term. Hence, it can be assumed to be zero in the calculation of the second order action, leading to $X^\mu = \delta^\mu_i X^i$.

In the comoving gauge, for which $\delta\varphi = 0$, the spatial metric have the general form
\begin{align}
	s_{ij} = a^2\left[\left(1 - 2\CR\right) \gamma_{ij} + 2D_i D_j E\right], \label{eq_gauge_gamma_ij}
\end{align}
where the choice of $E$ changes the shift but not the foliation, hence the gauge remains comoving. The choice $E = -\CC$ leads to $X^i = 0$ because the gauge invariant variable $\CC$ is defined as $\CC \coloneqq \CX - E$, where $X^i \eqqcolon D^i\CX$ \cite{2023_Vigneron_et_al_b}. The choice $E=0$ is the standard one, and the easiest in which calculating the second order action. For this choice, we have $X^\mu = \delta^\mu_i D^i\CC$, and the first and second orders of the reference curvature take the form
\begin{align}
    \overset{(1)}{\bar R}_{\mu\nu} &= 4K\delta_{(\mu}^0 \delta_{\nu)}^i \partial_i\CC' + 4K\delta_\mu^i \delta_\nu^j\,  D_i D_j \CC, \\
    \overset{(2)}{\bar R}_{\mu\nu} &= 2K\delta^0_\mu \delta^0_\nu\, D_k \CC' D^k\CC'  + 2K\left(D^k \CC D_k D_i D_j \CC + 2D_kD_i \CC D^k D_j\CC\right)\delta^i_\mu \delta^j_\nu + 2\overset{(2)}{\bar R}_{0i} \delta^0_{(\mu} \delta^i_{\nu)},
\end{align}
where $\overset{(2)}{\bar R}_{0i}$ is not needed for the calculation of the second order action.

We define the perturbations of the lapse and shift respectively as
\begin{align}
    N = a\left(1 + \alpha\right) \quad ; \quad N_i = a^2 \partial_i \beta.
\end{align}
The variation of the first order action~\eqref{eq_Action_ADM} with respect to $\alpha$ and $\beta$ gives the constraints equations, respectively as
\begin{align}
    \alpha = -\frac{\tilde\CR'}{\CH}, \quad\quad \Delta\left(\beta + \CC'\right) = -3K\CC' + \frac{1}{\CH} \Delta \tilde\CR - \epsilon \tilde\CR',
\end{align}
which also correspond, respectively, to equations~\eqref{eq_blind_pert_S_1} and~\eqref{eq_blind_pert_S_2}. From there, the goal is to calculate the action~\eqref{eq_Action_ADM} at second order in $\tilde\CR$ and $\CC$. The gauge choice~\eqref{eq_gauge_gamma_ij}, with $E=0$, implies
\begin{align}
    a^2 s^{ij} &= \gamma^{ij} +  \left[2\CR \gamma^{ij}\right] + \left[4\CR^2 \gamma^{ij}\right], \\
    \overset{(1)}{\Gamma}{}^a_{bc} &= -2\delta^a_{(b} D_{c)} \CR + \gamma_{bc} D^a \CR, \\
    \frac{1}{a^3}\sqrt{{\rm det} \, s_{ij}} &= 1 + \left[- 3\CR\right] + \left[\frac{3}{2}\CR^2\right].
\end{align}
Using $\alpha = -\frac{\tilde\CR'}{\CH}$, the background expansion laws and several integration by parts, we obtain
\begin{align}
    &\int \dd\tau \dd x^3 \sqrt{s}\left\{\frac{1}{N}\left(E^i{}_j E^j{}_i - E^2\right) + \left[\frac{1}{N}\left(\phi' + N^iD_i\phi\right)^2 + N D_i \phi D^i \phi - 2N V\right] + \frac{1}{N} N^i N^j \bar R_{ij} \right.\nonumber\\
    &\qquad \left.- \frac{2}{N} N^i\bar R_{i 0} - \frac{1}{N} \bar R_{00}\right\} = \int \dd\tau \dd x^3 a^2\sqrt{\gamma}\left[2\epsilon \left(\tilde\CR'\right)^2 - 6\left(K\CC'\right)^2 + 2K D_k \CC' D^k \CC' \right], \label{eq_EE}\\
    &\int \dd\tau \dd x^3 \sqrt{s} \, N \left(\CR^{(3)}_{ij} - \bar R_{ij}\right) s^{ij} = \int \dd\tau \dd x^3 a^2\sqrt{\gamma}\big[-2\epsilon D_k\tilde\CR D^k\tilde\CR \nonumber \\
    				&\qquad\qquad\qquad\qquad\qquad\qquad\qquad\qquad\qquad\qquad\quad\  \ + 6K^2 D_k \CC D^k\CC -2K\Delta\CC \Delta \CC\big].
\end{align}
Finally, the second order action for scalar modes takes the form
\begin{align}
S[\tilde\CR, \CC] &= \frac{1}{2}\int \dd\tau \dd x^3 2 a^2 m_{\rm p}^2 \sqrt{\gamma}\left\{
	 \epsilon\left[\left(\tilde\CR'\right)^2 - \left(D_k \tilde\CR \right)^2 \right] \right.\nonumber \\
	 &\qquad \left.- \left[3\left(K\CC'\right)^2 - 3\left(K D_k \CC \right)^2 + K D_k \CC' D^k \CC' - K\Delta\CC \Delta \CC \right] \right\}, \nonumber 
\end{align}
which, with some additional integration by parts, can be rewritten as
\begin{align}
S[\tilde\CR, \CC] = \frac{1}{2}\int \dd\tau \dd x^3 \sqrt{\gamma}\left(a m_{\rm p} \sqrt{2}\right)^2&\left\{\epsilon
	 \left[\left(\tilde\CR'\right)^2 - \left(D_k \tilde\CR \right)^2 \right] \right.\nonumber\\
	 &\ - K^2\left[\CC'\left(\Delta/K + 3\right)\CC' + \CC\Delta \left(\Delta/K + 3\right)\CC\right] \Big\}.
\end{align}
Introducing $\tilde v \coloneqq a m_{\rm p} \sqrt{2\epsilon} \, \tilde\CR$ and $C \coloneqq a m_{\rm p} K \sqrt{2} \, C$, we obtain the first two terms in the action~\eqref{eq_Action_Second_topo}. 
Interestingly, when deriving~\eqref{eq_EE}, the relation for $\Delta(\beta+\CC')$ is not needed as this term disappears during the calculation. On the contrary, in GR for $K\not=0$, $\beta$ remains present in the action, even without a Laplacian. This requires inverting the equation for $\Delta\beta$, leading to the presence of a non-local operator in the second order action and the field equation for $\CR$, as shown in \cite{2019_Handley}. The fact that this inversion of $\Delta\beta$ is not required in topo-GR is the reason why no non-local operator appears in the second order action.\saut

For tensor modes, the derivation in the framework of topo-GR is equivalent to the one in GR because the reference curvature up to second order does not feature tensor modes. The only differences come from the expansion laws not featuring the background curvature and the presence of $\overset{(0)}{R}_{\mu\nu} = 2K\gamma_{ij}$. The consequence is the replacement of $2K$ with $6K$ in the action, leading to
\begin{align}
S[h_{ij}] &= \frac{1}{2}\int \dd\tau \dd x^3 \sqrt{\gamma}\left(\frac{a m_{\rm p}}{2}\right)^2 \left[\left(h'_{ij}\right) ^2 - \left(D_k h_{ij}\right)^2 - 6K \left(h_{ij}\right)^2\right].
\end{align}
Introducing $\mathfrak h_{ij} \coloneqq a m_{\rm p} \frac{1}{2} h_{ij}$, we obtain the third term in the action~\eqref{eq_Action_Second_topo}.

\section{Canonical quantization of $C$}

The Lagrangian of $C$ features spatial derivatives of the velocity field $C'$ which cannot be removed with integration by parts. Therefore, to calculate the conjugate momentum of the field $C$, we need to come back to the general definition of the canonical momentum.

We consider a general Lagrangian depending on a field $q$, its time derivative $\dot q$ and the spatial derivatives (first order or higher) of $q$ and $\dot q$, i.e.  $\CL\left(q, D_i q, D_iD_jq, ..., \dot q, D_i \dot q, D_i D_j \dot q, ...\right)$. The conjugate momentum~$\pi$ of~$q$ is defined as the functional derivative of $\CL$ with respect to $\dot q$, i.e.  for a trial function $\phi$ we have
\begin{align} \label{eq_momentum_def}
    \int \dd x^3 \pi \, \phi
        &\coloneqq \int \dd x^3 \frac{\delta \CL}{\delta \dot q} \, \phi \nonumber \\
        &:= \int \dd x^3 \left[\partial_{\dot q} \CL + \frac{\partial\CL}{D_i \dot q} D_ i + \frac{\partial\CL}{D_iD_j \dot q} D_iD_j + ...\right] \, \phi,
\end{align}
which becomes, after integration by parts,
\begin{align}
     \int \dd x^3 \, \pi \, \phi&= \int \dd x^3  \left[\partial_{\dot q} \CL - D_i \frac{\partial\CL}{D_i \dot q} + D_iD_j \frac{\partial\CL}{D_iD_j \dot q} + ...\right] \phi .
\end{align}
Up to second spatial derivatives, this yields
\begin{align} \label{eq_momentum_def_2}
    \pi \coloneqq \frac{\partial \CL}{\partial \dot q} - D_i \frac{\partial\CL}{\partial {D_i \dot q}} + D_i D_j \frac{\partial\CL}{\partial {D_i D_j \dot q}} + ...
\end{align}
Due to the dependence of $\CL$ in $D_i\dot q$, we get extra terms in the definition of $\pi$, compared to the usual formula $\pi = \partial\CL/\partial{\dot q}$. Eq~\eqref{eq_momentum_def} also shows that adding a boundary term in the Lagrangian, e.g. $\CL \rightarrow \CL + \Delta \dot q^2$, does not change the expression for the momentum.

Now using Eq~\eqref{eq_momentum_def_2} for the Lagrangian of $C$, in the limit $\tau \rightarrow -\infty$, so that we neglect $a''/a$, i.e.
\begin{align}
    \CL_C \coloneqq -\frac{1}{2}\left(3C'^2 + \frac{1}{K} C'\Delta C' + 3C\Delta C + \frac{1}{K}C \Delta\Delta C\right),
\end{align}
the conjugate momentum of $C$ is
\begin{align}
    \pi_C = -\left(\Delta/K + 3\right) C'.
\end{align}
Performing a canonical quantization of $C$ and considering the minimal energy state of the Hamiltonian as the initial condition, we constrain $C$ to a single initial oscillating mode 
\begin{align}
    C \ \overset{\tau \rightarrow - \infty }{\propto} \ e^{\pm ik\tau}, \qquad k^2 \not= 3K.
\end{align}
From the canonical commutation rule $[\hat C,\hat \pi_C] = i\hbar$, we select the mode $e^{-ik\tau}$ and constrain the amplitude such that
\begin{align}
    C \overset{\tau \rightarrow - \infty }{\sim} \left(\frac{K}{k^2-3K}\right)\frac{1}{\sqrt{2k}} e^{-ik\tau}, \quad k^2 > 3K. \label{eq_IC_C_bis}
\end{align}
The mode $k^2 = 3K$ is not physical since $C$ disappears from the action for that mode. Additionally, for $K>0$, the wave-number $k$ is discrete and given by ${k^2 = n(n+2)K}$ for $n \in \mathbb N$. Consequently, the first physical mode is $k^2 \geq 8K$, implying that the initial condition~\eqref{eq_IC_C_bis} is well-defined for all physical modes and the amplitude of $C$ is bounded from above by a finite value, for $K$ positive or negative.

\end{document}